\documentclass[12pt]{article}
\usepackage{graphicx,setspace,fullpage, amsmath,bm} %
\usepackage{amsthm,amsmath,amsfonts,amssymb,bm,graphicx}
\usepackage{natbib}
\newcommand\scalemath[2]{\scalebox{#1}{\mbox{\ensuremath{\displaystyle #2}}}}

\title{Bayesian Variable Selection in Multivariate Regression Under Collinearity in the Design Matrix}

\author{Joyee Ghosh\footnote{Joyee Ghosh is Associate Professor, Department of Statistics and Actuarial Science, The University of Iowa.} \ and Xun Li\footnote{Xun Li is a 
PhD graduate from Department of Statistics and Actuarial Science, The University of Iowa. She is now Lead Modeler at Chime.}}

\date{}

\begin{document}

\maketitle

\begin{center}
 {Abstract}   
\end{center}
We consider the problem of variable selection in Bayesian multivariate linear regression models, involving multiple response and predictor variables, under multivariate normal errors. 
In the absence of a known covariance structure, specifying a model with a non-diagonal covariance matrix is appealing. Modeling dependency in the random errors through a non-diagonal covariance matrix is generally expected to lead to improved estimation of the regression coefficients. In this article, we highlight an interesting exception: modeling the dependency in errors can significantly worsen both estimation and prediction. We demonstrate that Bayesian multi-outcome regression models using several popular variable selection priors can suffer from poor estimation properties in low-information settings--such as scenarios with weak signals, high correlation among predictors and responses, and small sample sizes. In such cases, the simultaneous estimation of all unknown parameters in the model becomes difficult when using a non-diagonal covariance matrix.
Through simulation studies and a dataset with measurements from NIR spectroscopy, we illustrate that a two-step procedure--estimating the mean and the covariance matrix separately--can provide more accurate estimates in such cases. Thus, a potential solution to avoid the problem altogether is to routinely perform an additional analysis with a diagonal covariance matrix, even if the errors are expected to be correlated. 

\doublespacing

\section{Introduction}
There is a rich literature on methods for multi-outcome regression models, also referred to as multivariate regression. A paper by \citet{Brei:Frie:1997} 
proposed a method called curds and whey for multivariate regression. This paper was one of the first to illustrate the gain in prediction with multivariate regression models over univariate ones when the errors are correlated. The method utilized cross-validation to construct a linear combination of ordinary least squares estimates for univariate regression. In the Bayesian paradigm, a seminal paper by \citet{Brow:Vann:Fear:1998} introduced variable selection with spike-and-slab priors \citep{Geor:Mccu:1993, Geor:McCu:1997} for multivariate regression with a non-diagonal covariance matrix. In this paper, each predictor variable was assumed to be associated with all the response variables or none of them, with spike-and-slab priors. As Bayesian variable selection evolved, many more generalizations of the models, priors, and algorithms have been proposed for multivariate regression models, such as those based on sparse seemingly unrelated regression models \citep{Wang:2010}, Gaussian graphical models \citep{Bhad:Mall:2013}, and reduced rank regression \citep{Chak:Bhat:Mall:2019}.

In this article our focus will be on the classical form of the multivariate regression model with normal errors and variable selection priors. In that setup, \citet{Desh:Rock:2019} used a more general version of the spike-and-slab priors than \citet{Brow:Vann:Fear:1998}, where a predictor variable could be important for a response variable but unimportant for another response variable. They used continuous spike-and-slab lasso priors for variable selection and covariance selection. Their objective was to estimate the posterior mode with an Expectation Conditional Maximization (ECM) algorithm.
\citet{Bott:Bant:2021} used spike-and-slab priors where the spike corresponded to a point mass prior on the regression coefficient that set it exactly to zero. The authors used evolutionary Monte Carlo for full posterior exploration. \citet{Li:Ghos:Vill:2023:tas} used 
spike-and-slab priors of a similar nature to compare predictions under models with a non-diagonal versus a diagonal covariance matrix in a moderate dimensional model space. Their results suggested that the improvement in estimation of regression coefficients with a non-diagonal covariance matrix could be substantial when the errors have high correlation. There was also an accompanying improvement in prediction, but the predictive gain was relatively small.  

Spike-and-slab priors are usually considered ideal for variable selection, especially those that allow the regression coefficients to be exactly zero. However, the computation associated with these priors does not scale well
with high dimensions, when the posterior becomes highly multimodal. As a result, a growing body of literature on Bayesian variable selection now utilizes the popular horseshoe priors \citep{Carv:Pols:Scot:2010, Bhad:Datt:Pols:Will:2017} that act as a good proxy for spike-and-slab priors \citep{Datt:Ghos:2013}. These are continuous global local shrinkage priors \citep{Pols:Scot:2010} under which posterior computation can be more amenable. \citet{Bai:Ghos:2018} extended the idea of global local shrinkage priors for univariate regression to the multivariate setup and established sufficient conditions for posterior consistency of the regression coefficient matrix when the covariance matrix is known. Their prior includes the popular horseshoe prior as a special case, and a Gibbs sampler can be used for posterior computation. \citet{Kund:Mitr:Gask:2021}
proposed an alternative form of a global local shrinkage prior
and also established conditions for posterior consistency.

The organization of this article is as follows. In Section 
\ref{Sec:problem} we first use a single simulated dataset to illustrate the problem of estimation with a non-diagonal covariance matrix when there is relatively less information in the data due to collinearity in the design matrix. We show that a simple two-step approach to estimation can provide better results under this setting.
In Section \ref{Sec:simstudy} we conduct a more comprehensive simulation study with more replications and include comparisons with several Bayesian variable selection methods that use a non-diagonal covariance matrix. In Section \ref{Sec:real} we apply the methods to a dataset with measurements from NIR spectroscopy which has high correlation in the design matrix. Finally, in 
Section \ref{Sec:Disc} we summarize our findings and provide some recommendations.

\section{Low Information Under Collinearity: Potential Problem and a Possible Solution} \label{Sec:problem}
It is generally considered desirable to jointly estimate the mean and the non-diagonal covariance matrix in Bayesian variable selection for multivariate regression. Some authors have mentioned the corresponding gain in estimation and prediction, especially when regression coefficients are small and there is a tendency for a common pattern of importance of predictors across response variables, which helps in borrowing information across response variables and can boost performance \citep{Desh:Rock:2019, Li:Ghos:Vill:2023:tas}. However, in certain cases, the information in the data may be limited due to a small sample size, collinearity in the predictors and random errors, and a relatively small magnitude of the regression coefficients. We first set up the notation for the model and then illustrate with an example that simultaneous estimation of the mean and covariance matrix may not perform well under such circumstances. 

\subsection{Multivariate Normal Model With a Non-diagonal Covariance Matrix and Spike-and-slab Priors for Regression Coefficients}
\label{subsec:mvtmodel}
Let $\bm{y}_i$ and $\bm{x}_i$ denote the $q \times 1$ and $p \times 1$ vectors of the response and predictor variables, respectively, for the $i$th observation, where $i=1,2,\dots,n$. We assume that the response and predictor variables have been centered and scaled to have mean 0 and standard deviation 1. The multivariate regression model is given as follows:
\begin{equation}
\bm{y}_i = \bm{B}'\bm{x}_i  + \bm{\epsilon}_i, \ i=1,2,\dots, n, 
\label{eq:mvtnormal.vec}
\end{equation}
where $\bm{B}$ is the $p \times q$ matrix of regression coefficients, 
and $\bm{\epsilon}_i$ is the $q \times 1$ vector of random errors for observation $i$. We assume $\bm{\epsilon}_i | \bm{\Sigma} \stackrel{iid}{\sim} {\rm N}_q(\bm{0}, \bm{\Sigma}), \ i=1,2,\dots,n$, where ${\rm N}_q(\bm{0}, \bm{\Sigma)}$ is the $q$-dimensional multivariate normal distribution with mean vector $\bm{0}$ and covariance matrix 
$\bm{\Sigma}$. 

We fit the model in (\ref{eq:mvtnormal.vec}) with the following priors. The prior on $\bm{\Sigma}$ is taken to be inverse Wishart with degrees of freedom and scale matrix $q+2$ and $0.5\bm{I}$, where $\bm{I}$ is the $q \times q$ identity matrix. This particular choice of the degrees of freedom parameter ensures the existence of a prior mean of $\bm{\Sigma}$ which is given by $0.5\bm{I}$. Since the response variables are standardized to have variance 1, the values of error variance cannot exceed 1. The choice  $0.5\bm{I}$ assigns a prior mean error variance of 0.5, which seems a reasonable prior guess when no other information on the error variance is available. Let $\gamma_{jk}=1$ when the $j$th predictor is associated with the $k$th response variable and $\gamma_{jk}=0$ otherwise. 
Then for $j=1,2,\dots,p$ and $k=1,2,\dots,q$,  
\begin{eqnarray}
    \gamma_{jk} | \pi  &\stackrel{iid}{\sim} & 
    \mathrm{Bernoulli}(\pi), \label{eq:gammaprior} \\
  \bm{B}_{jk} | \gamma_{jk}  &\stackrel{ind}{\sim}& \gamma_{jk} \mathrm{N}(0,1) +(1-\gamma_{jk}) \delta_{\{0\}},
  \label{eq:betaprior.i.normal}
\end{eqnarray}
where $ind$ is an abbreviation for independently, $\mathrm{N}(0,1)$ denotes the univariate Normal distribution with mean 0 and variance 1 corresponding to the slab for active/important predictors, and $\delta_{\{0\}}$
denotes a point mass at 0 corresponding to the spike for inactive/noise predictors. The prior variance of 1 for the normal prior is chosen as the predictors have been standardized. Here we set $\pi=0.5$ which corresponds to a discrete uniform prior over the space of models. This prior choice corresponds to the "\textit{dep-data-ind-prior spike-and-slab}" of \citet{Li:Ghos:Vill:2023:tas} which refers to dependent errors via the non-diagonal $\bm{\Sigma}$ and independent spike-and-slab priors on the regression parameters. Alternatively, a Beta prior can be assigned to $\pi$, especially for larger model spaces. 

\subsection{An Example to Illustrate the Problem Under Collinearity}
\label{subsec:example}
We take $p=10$ and $q=4$. We generate the $p \times 1$ predictors independently for $i=1,2,\dots, n$, from a multivariate normal distribution with a $p \times 1$ mean vector $\bm{0}$ and a $p \times p$ covariance matrix whose $(i,j)$th element is $0.7^{|i-j|}$ to have moderate correlation in the design matrix. The $q \times q$ error covariance matrix has 1 on the diagonals and 0.9 on off-diagonals to represent high correlation among the $q$ random errors. The first and second rows of the regression coefficient matrix $\bm{B}$ are set as $(1.25, 1.25, 1.25, 1.25)$ and $(-1, -1, -1, -1)$ respectively. The remaining rows are all equal to $(0, 0, 0, 0)$. This implies the true model is sparse, with only two important predictor variables that share the same relationship with all the response variables. 

We generate $n=40$ samples from model (\ref{eq:mvtnormal.vec}) with the aforementioned parameters. To analyze the data under the specified prior, we run a Markov chain Monte Carlo (MCMC) algorithm for 10,000 iterations with a burn-in of 100 samples, and given below is the corresponding MCMC estimate of the posterior mean. We find that the posterior mean shrinks the non-zero regression coefficients in the first two rows to nearly zero.
\begin{equation}
\scalemath{0.87}{
\hat{E}(\bm{B} | \bm{y}_1,\dots,\bm{y}_{40}) = \left(
 \begin{array}{rrrr}
 -0.005 & 0.002 & -0.001 & 0.013 \\ 
  -0.007 & -0.001 & -0.001 & 0.003 \\ 
   -0.003 & 0.000 & 0.000 & -0.005 \\ 
   -0.012 & -0.001 & 0.002 & 0.000 \\ 
  -0.021 & -0.038 & 0.015 & 0.005 \\ 
   -0.014 & -0.001 & 0.007 & 0.001 \\ 
   -0.005 & 0.002 & -0.001 & 0.001 \\ 
   -0.001 & 0.007 & -0.001 & -0.003 \\ 
  0.003 & -0.011 & 0.001 & 0.018 \\ 
   -0.002 & -0.009 & 0.020 & 0.000 \\ 
 \end{array}\right).
 }
 \label{eq:mvt.beta.est.n40}
 \end{equation}

We now generate a sample of size $n=200$ while keeping everything else the same. The 
estimated posterior mean is as follows:
\begin{equation}
\scalemath{0.87}{
\hat{E}(\bm{B} | \bm{y}_1,\dots,\bm{y}_{200}) = \left(
 \begin{array}{rrrr}
  1.233 & 1.194 & 1.242 & 1.252 \\ 
   -1.002 & -1.018 & -1.003 & -1.065 \\ 
  0.000 & 0.001 & -0.005 & 0.001 \\ 
   -0.004 & 0.006 & 0.000 & -0.011 \\ 
   -0.001 & -0.001 & -0.001 & -0.002 \\ 
  0.009 & -0.030 & -0.020 & -0.006 \\ 
   0.001 & -0.002 & -0.001 & 0.000 \\ 
   0.000 & 0.000 & -0.001 & 0.000 \\ 
  0.000 & 0.000 & 0.000 & -0.001 \\ 
   0.000 & 0.000 & 0.001 & -0.001
 \end{array}
 \right).}
\end{equation}
The above result shows that the estimate can be quite close to the true value of $\bm{B}$ when the sample size is relatively large. This suggests that when the information in the data is limited, the estimates obtained under a model with a non-diagonal $\bm{\Sigma}$ and 
spike-and-slab priors for $\bm{B}$ can sometimes be affected in an extreme manner.

\subsection{A Potential Solution With a Two-step Procedure}
\label{subsec:twostep}
Estimating the mean and non-diagonal covariance matrix simultaneously may result in inaccurate estimates of the parameters under spike-and-slab priors. A possible solution is to: 
\begin{enumerate} 
\item Estimate the mean using separate univariate regression models with spike-and-slab priors. 
\item Calculate the residuals under the fitted univariate models and use them to estimate the non-diagonal $\bm{\Sigma}$. 
\end{enumerate}
Note that we are not advocating the use of the aforementioned simpler, misspecified model (when the true $\bm{\Sigma}$ is non-diagonal) to replace the more general model with a non-diagonal $\bm{\Sigma}$. 
However, we suggest that routinely performing the analysis under both the simpler and the more general model can help mitigate issues related to low-information data and reduced estimation accuracy.

In \textbf{Step 1}, we consider model (\ref{eq:mvtnormal.vec}) with the additional constraint that $\bm{\Sigma}={\rm diag}(\sigma_1^2,\dots,\sigma_q^2)$, a $q \times q$ diagonal matrix. We specify the same prior on 
$\bm{B}$ as in Section \ref{subsec:mvtmodel} and let $1/\sigma^2_k \stackrel{iid}{\sim}$ Gamma$(1.5, 0.25), \ k=1,2,\dots,q$, with 1.5 and 0.25 being the shape and rate parameters, respectively. This choice 
is consistent with the univariate version of the inverse Wishart 
prior on the non-diagonal $\bm{\Sigma}$ in Section \ref{subsec:mvtmodel}. This prior was called "\textit{ind-data-ind-prior spike-and-slab}" by \citet{Li:Ghos:Vill:2023:tas} which refers to independent errors because of the diagonal $\bm{\Sigma}$ and independent spike-and-slab priors on the regression parameters. The estimated posterior mean of $\bm{B}$ under this model and prior is given below in (\ref{eq:twostep.beta.est}). While the estimate is not in perfect agreement with the true value of $\bm{B}$, it is immediately clear that the first two rows corresponding to the active predictors are much more accurately estimated in this model compared to the estimates in (\ref{eq:mvt.beta.est.n40}) under the multivariate model with a non-diagonal $\bm{\Sigma}$.
\begin{eqnarray}
  \hat{E}^{two-step}(\bm{B} | \bm{y}_1,\dots,\bm{y}_{40}) = 
  \left(
 \begin{array}{rrrr}
  0.853 & 0.973 & 0.947 & 1.032 \\ 
   -0.921 & -1.019 & -1.011 & -1.006 \\ 
   -0.005 & 0.003 & 0.009 & -0.014 \\ 
   -0.048 & -0.037 & -0.030 & -0.040 \\ 
   -0.009 & -0.019 & 0.008 & 0.009 \\ 
   0.032 & 0.081 & 0.152 & 0.078 \\ 
  -0.023 & -0.021 & -0.044 & -0.033 \\ 
   0.005 & 0.022 & 0.008 & -0.009 \\ 
   0.092 & 0.040 & 0.077 & 0.175 \\ 
   0.002 & 0.002 & 0.029 & -0.001 \\ 
 \end{array}\right).
\label{eq:twostep.beta.est}
 \end{eqnarray}

 If one is also interested in estimating the non-diagonal covariance matrix $\bm{\Sigma}$, we outline a simple approach for doing so in \textbf{Step 2}, as follows. We estimate the residuals, say $\bm{e}_i$ as $\bm{e}_i = \bm{y}_i - \hat{E}^{two-step}(\bm{B} | \bm{y}_1,\dots,\bm{y}_{n})'\bm{x}_i$. We assume $\bm{e}_i \stackrel{iid}{\sim}$ N$_q(\bm{0},\bm{\Sigma}), \  i=1,2,\dots,n$. Actually, the unobserved random errors $\bm{\epsilon}_i$s are independently and identically distributed as N$_q(\bm{0},\bm{\Sigma})$, whereas $\bm{e}_i$s are their point estimates. This simplified assumption about $\bm{e}_i$s yields a closed form posterior distribution for $\bm{\Sigma}$ under a conjugate prior. Using an inverse Wishart prior for $\bm{\Sigma}$ with the degrees of freedom and the scale matrix as $q+2$ and  $0.5\bm{I}$ respectively, the posterior is also inverse Wishart with the corresponding parameters being $q+2+n$ and $0.5\bm{I} + S_n$, where
$S_n= \sum_{i=1}^n \bm{e}_i\bm{e}_i'$. We estimate $\bm{\Sigma}$ by the 
posterior mean, given by 
$\frac{0.5\bm{I} + S_n}{n+1}$. 

In the next section, we carry out a simulation study to compare the above two-step approach with several popular Bayesian variable selection methods for multivariate regression with a non-diagonal $\bm{\Sigma}$. Specifically, we compare the methods with respect to the estimation of $\bm{B}$ and $\bm{\Sigma}$, and the prediction of new response variables generated from the true model.

\section{Simulation Study} \label{Sec:simstudy}
We generate 25 simulated datasets using the same setup described in Section \ref{subsec:example} to enable a more comprehensive study of the problem. We compare the following five methods, all of which, except the two-step approach, use a non-diagonal $\bm{\Sigma}$. 

\begin{enumerate}
\item The \textbf{two-step} approach: This method was described in Section \ref{subsec:twostep}.

\item The \textbf{d}ep-data-ind-prior \textbf{s}pike-and-\textbf{s}lab (\textbf{dss}): This procedure was described in Section \ref{subsec:mvtmodel}.

\item The \textbf{m}ultivariate \textbf{s}pike-and-\textbf{s}lab \textbf{l}asso (\textbf{mssl}): This method by \citet{Desh:Rock:2019} estimates the posterior mode corresponding to continuous spike-and-slab lasso (Laplace) priors. The authors consider the same model in (\ref{eq:mvtnormal.vec}) and the model space prior in (\ref{eq:gammaprior}). They consider a $\mathrm{Beta}(a_{\pi},b_{\pi})$ prior on 
$\pi$. For $\bm{B}$ they propose the following prior:
\begin{equation}
    \bm{B}_{jk} | \gamma_{jk}, \lambda_1, \lambda_0  \stackrel{ind}{\sim} \gamma_{jk} \phi_1(\lambda_1) +(1-\gamma_{jk}) \phi_0(\lambda_0),
  \label{eq:betaprior.sslasso}
\end{equation}
where $\phi_1(\lambda_1)$ and $\phi_0(\lambda_0)$ are Laplace distributions with location parameters 0 and scale parameters
$1/\lambda_1$ and $1/\lambda_0$, corresponding to the spike and slab prior distributions, respectively. For the diagonal elements of the precision matrix $\bm{\Sigma}^{-1}$, the authors use  exponential distributions as priors.
For the off-diagonals they propose spike-and-slab Laplace priors for shrinkage, when $q$ is large. In our implementation we only consider the slab part of the prior for the off-diagonals in the precision matrix, as $q$ is not very large in our examples.

\item The \textbf{m}ultivariate \textbf{B}ayesian model with \textbf{s}hrinkage \textbf{p}riors (\textbf{mbsp}): This approach of \citet{Bai:Ghos:2018} proposes global local shrinkage priors for variable selection.
The model is same as in (\ref{eq:mvtnormal.vec}) and the prior on $\bm{\Sigma}$ is inverse Wishart. Given the global shrinkage parameter $\tau$ and local shrinkage parameters $\psi_1,\dots,\psi_p$, 
they specify a matrix normal prior distribution for $\bm{B}$ with a $p \times q$ location matrix $\bm{0}$, and $p \times p$ and $q \times q$ scale matrices $\tau \mathrm{diag}(\psi_1,\dots,\psi_p)$ and $\bm{{\Sigma}}$, respectively.
Additionally $\psi_1,\dots,\psi_p$ are assigned a polynomial-tailed prior density of a particular form that leads to a family of global local shrinkage priors for $\bm{B}$. In particular, we consider the horseshoe prior which is a member of this family of priors. The authors recommend setting the global shrinkage parameter $\tau$ at 
$\frac{1}{p\sqrt{n \log(n)}}$ based on an asymptotic justification. 

\item The \textbf{d}ep-data-ind-prior \textbf{h}orse\textbf{s}hoe (\textbf{dhs}): This method specifies independent horseshoe priors given the global shrinkage parameters, and was implemented by \citet{Li:Ghos:Vill:2023:tas} and  \citet{Kund:Mitr:Gask:2021}. This was referred to as the Naive Horseshoe by \citet{Kund:Mitr:Gask:2021} due to its independent structure. This method considers the same model in (\ref{eq:mvtnormal.vec})
with an inverse Wishart prior on $\bm{\Sigma}$. Here 
$B_{jk} | \xi_{jk}, \tau_k \stackrel{ind}{\sim} \textrm{N}(0,\xi_{jk} \tau_k )$,
$\xi_{jk}^{\frac{1}{2}} \stackrel{iid}{\sim} C^+(0,1)$, and $\tau_{k}^{\frac{1}{2}} \stackrel{iid}{\sim} C^+(0,1)$, where 
$C^+(0,1)$ is the half-Cauchy distribution with location parameter 0 and scale parameter 1 \citep{Carv:Pols:Scot:2010}. The $\tau_k$s are the $q$ global shrinkage parameters, one for each response variable, and the $\xi_{jk}$s are the $pq$ local shrinkage parameters, one for each predictor-response combination.
\end{enumerate}

The methods two-step, dss, and dhs can be implemented with the JAGS code provided by \citet{Li:Ghos:Vill:2023:tas}. The other two methods can be implemented by R packages {\tt mSSL}
 \citep{mSSL} and {\tt MBSP} \citep{mbsp}. 
Except for mssl, all the other methods are based on MCMC, which are run for 10,000 iterations including a burn-in of 100, and the posterior mean is used as the MCMC estimate for parameters of interest. Trace plots and MCMC standard errors were used to determine an appropriate number of MCMC iterations. The maximum number of iterations in the ECM algorithm for mssl is also set at 10,000 iterations, and the estimated posterior mode is used in this case. The spike-and-slab priors in two-step and dss are based on a discrete uniform prior on the model space, that corresponds to $\pi=0.5$. The default choice for the inclusion probability $\pi$ in the R package {\tt mSSL} is a Beta$(1,pq)$ prior, which can shrink regression coefficients aggressively to zero. This could be appropriate in more high-dimensional settings; however, it is not ideal in this example, as the coefficients were shrunk even under the more lenient discrete uniform prior in Section \ref{subsec:example} in (\ref{eq:mvt.beta.est.n40}). So we use a Beta$(1,1)$ prior instead, which is closer to the discrete uniform prior than the Beta$(1,pq)$ prior. All other arguments for {\tt mSSL} and {\tt MBSP} are set at their default values provided in the two R packages.  

\begin{figure*}[htb]
    \centering
 \includegraphics[width=2.1in]{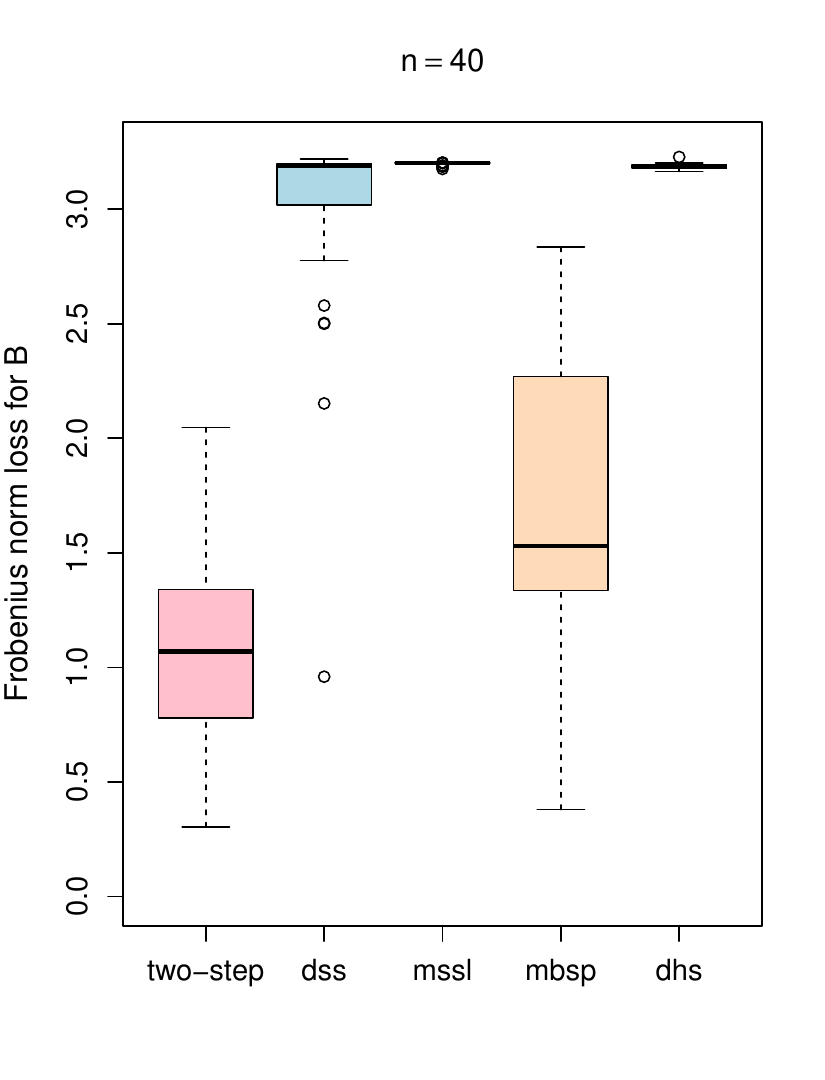}
    \includegraphics[width=2.1in]{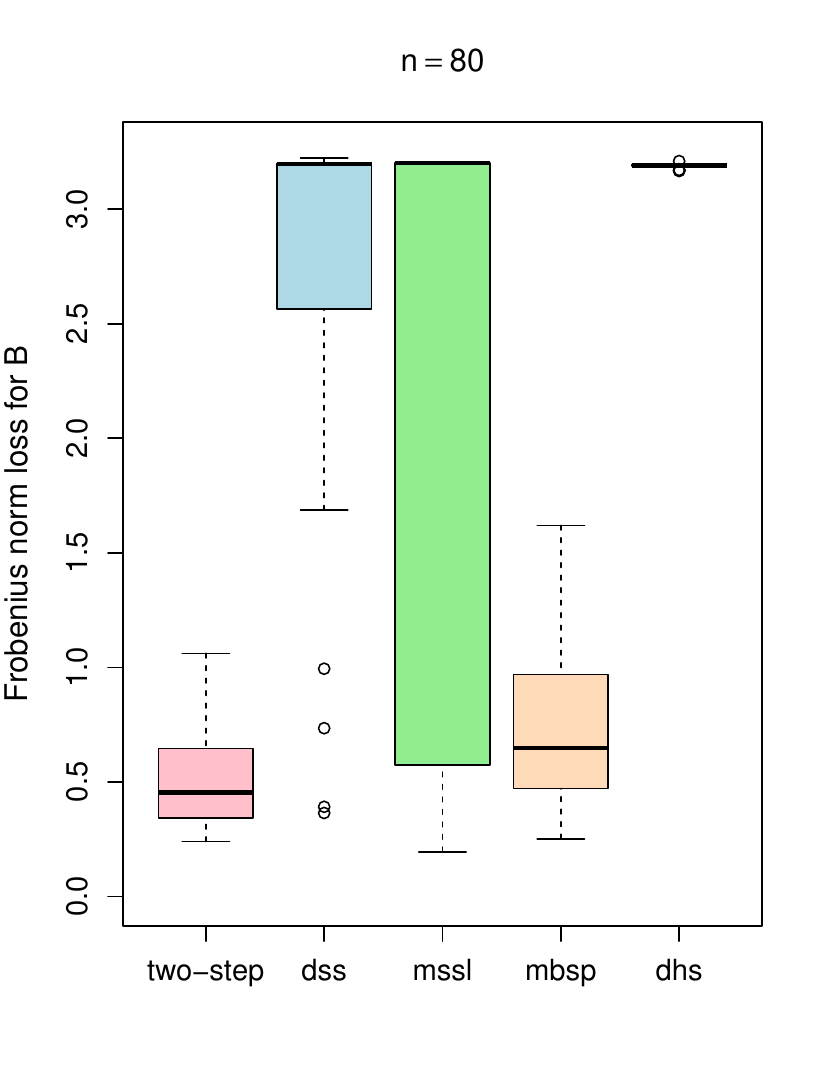}
    \includegraphics[width=2.1in]{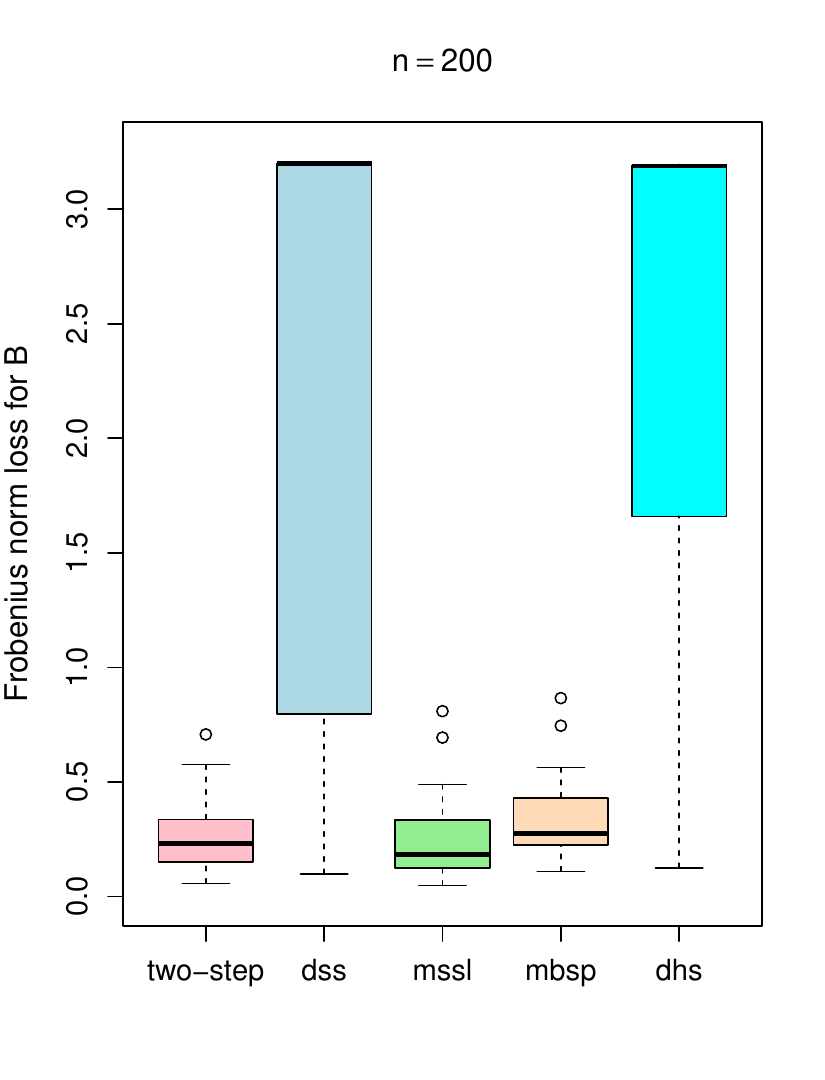}
    \\
     \includegraphics[width=2.1in]{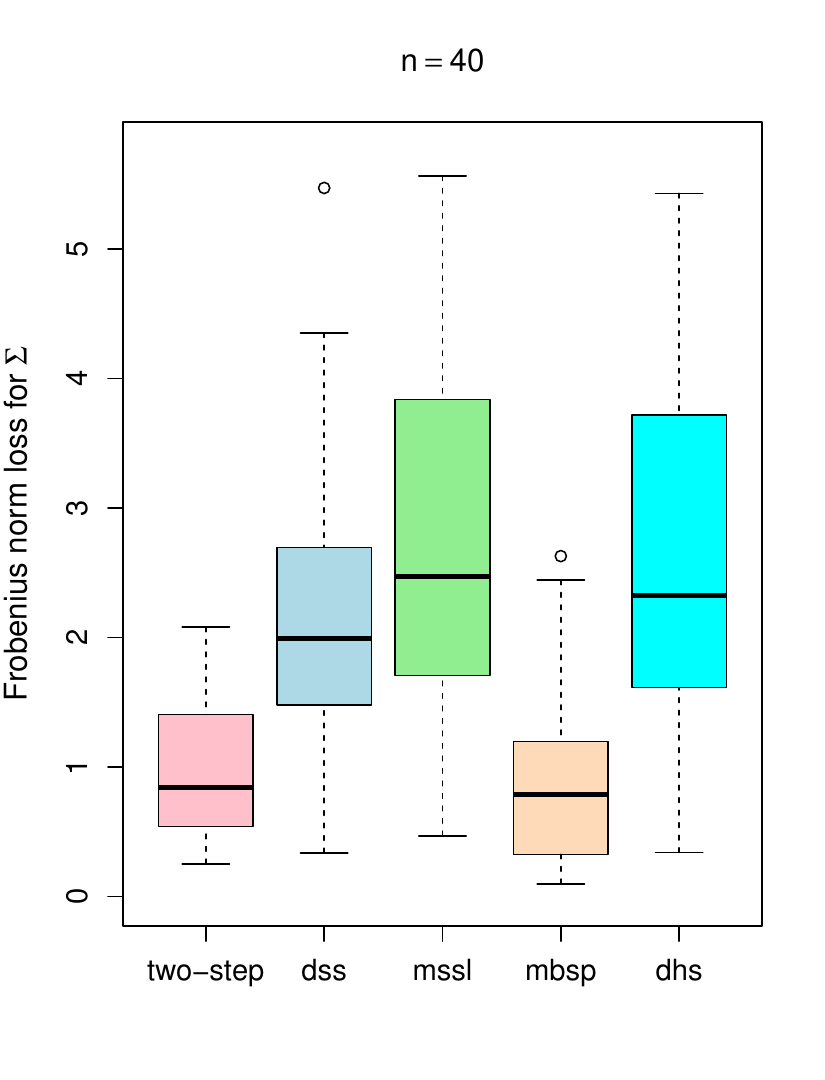}
    \includegraphics[width=2.1in]{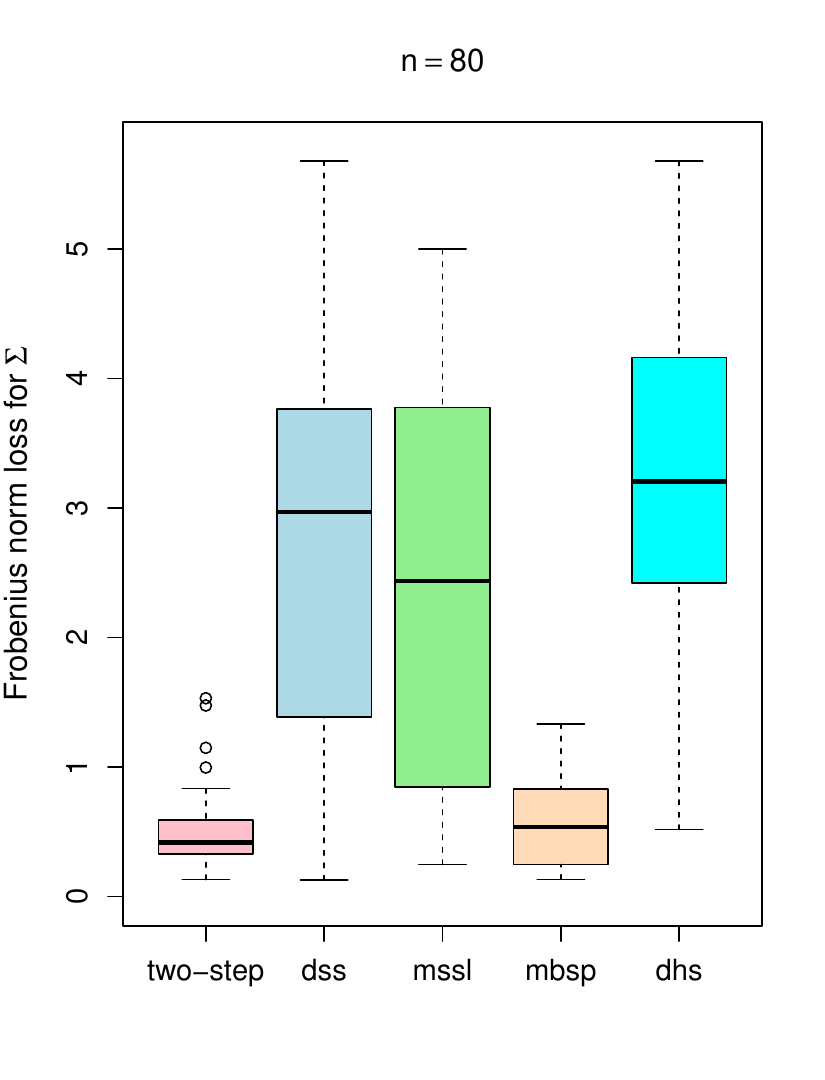}
    \includegraphics[width=2.1in]{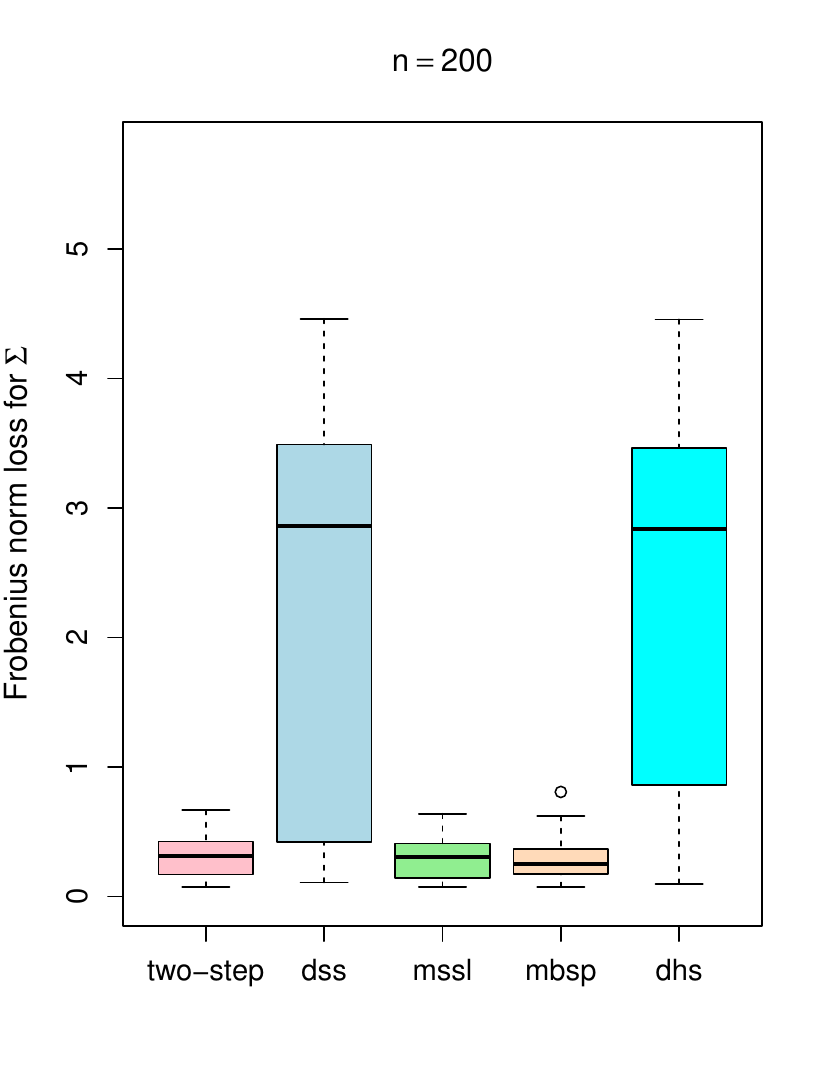}
    \\
     \includegraphics[width=2.1in]{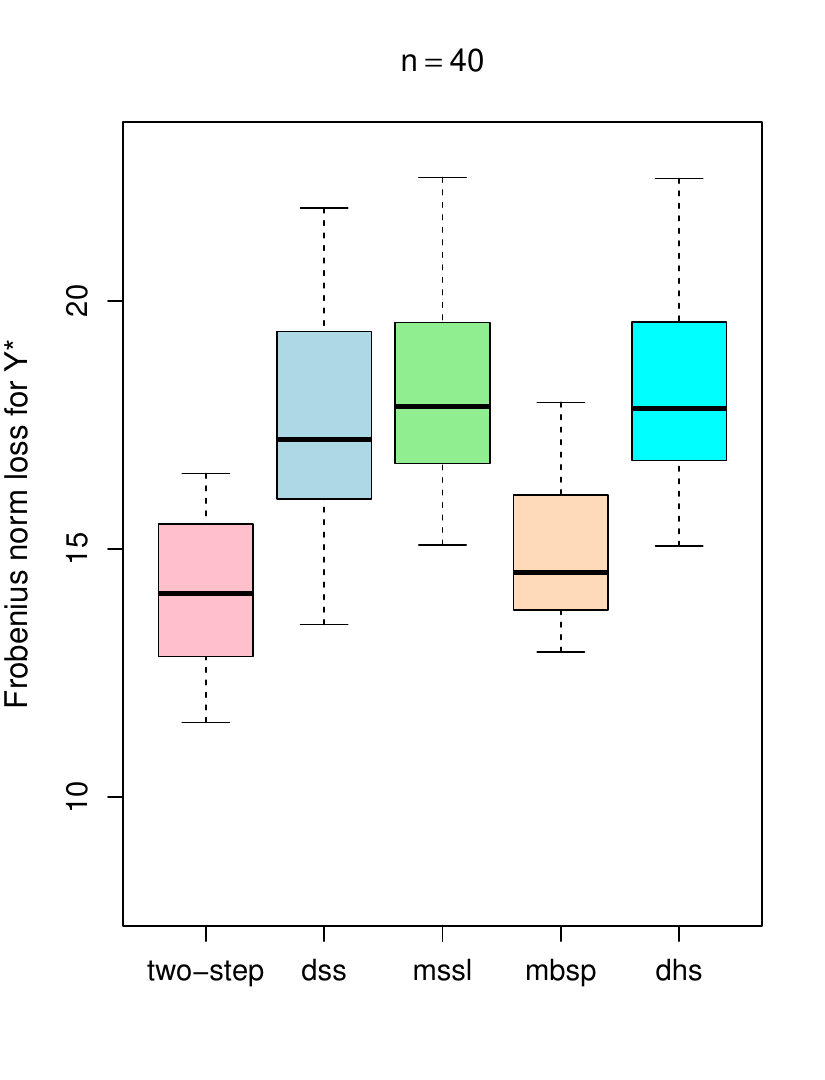}
    \includegraphics[width=2.1in]{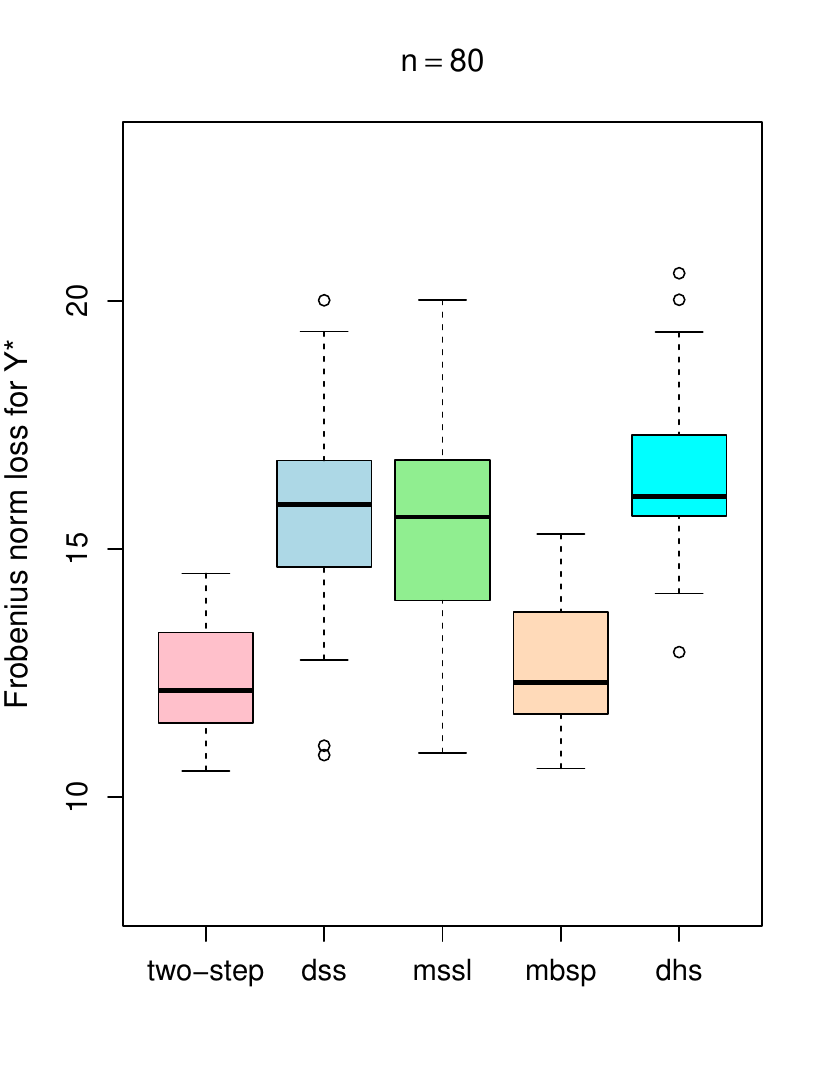}
    \includegraphics[width=2.1in]{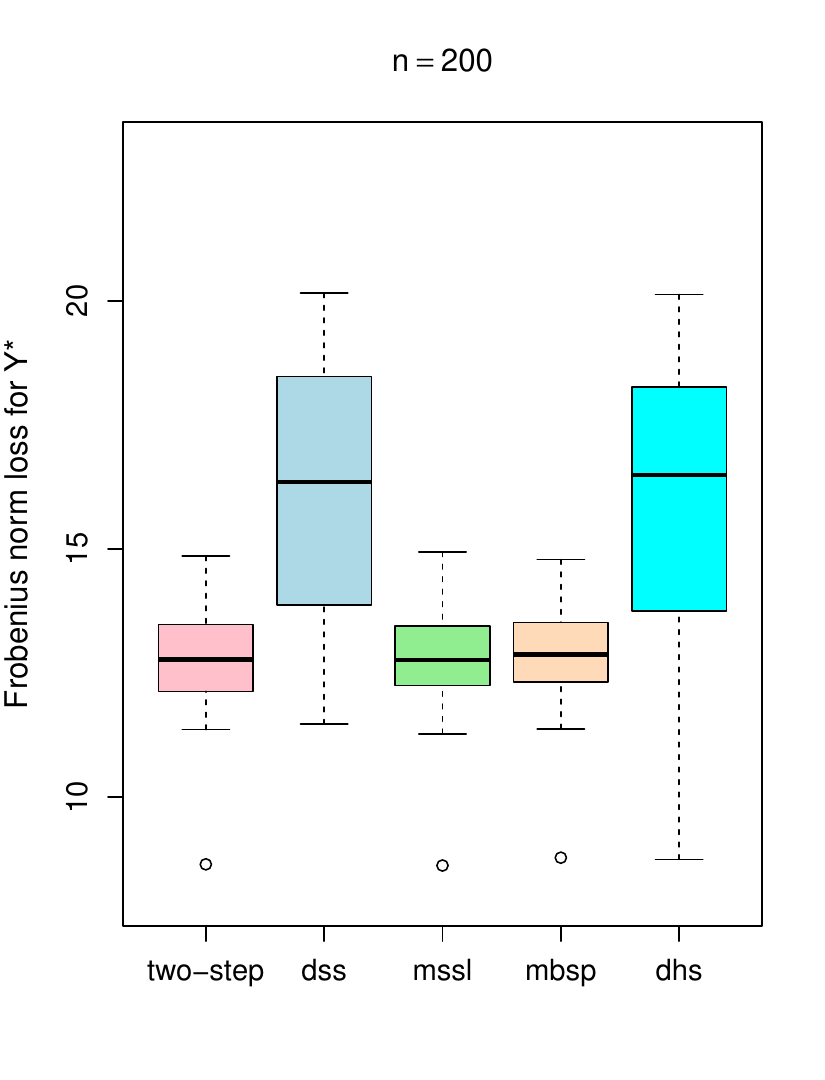}
    \caption{Box plots showing the values of Frobenius norm loss for estimating $\bm{B}$, $\bm{\Sigma}$, and $\bm{Y}^*$ for the 25 replicates in the simulation study in Section 4.}
    \label{fig:simstudy}
\end{figure*}

As in Section \ref{subsec:example}, we first use $n=40$ for the estimation of $\bm{B}$ and $\bm{\Sigma}$, and generate an additional sample of size $n^*=40$ as test data for prediction, denoted by the
$n^* \times q$ matrix of response variables
$\bm{Y}^*=(\bm{y}_1^*,\bm{y}_2^*,\dots,\bm{y}_{n^*}^*)'$ corresponding to the $n^* \times p$ design matrix $\bm{X}^*=(\bm{x}_1^*,\bm{x}_2^*,\dots,\bm{x}_{n^*}^*)'$. For each method, we report the Frobenius norm loss for estimating the matrices $\bm{B}$, $\bm{\Sigma}$, and $\bm{Y}^*$ by 
$||(\bm{B} - \hat{\bm{B}})||_{F}$, 
$||(\bm{\Sigma}- \hat{\bm{\Sigma}})||_{F}$, and $||(\bm{Y}^*-\hat{\bm{Y}}^*)||_{F} 
= ||(\bm{Y}^*-\bm{X}^*\hat{\bm{B}}^*)||_{F} $, respectively, where the Frobenius norm of an $I \times J$ matrix $\bm{A} = (a_{ij})$ is given by 
$||\bm{A}||_F=\sqrt{\sum_{i=1}^{I}\sum_{j=1}^{J} a_{ij}^2}$. Smaller values of the Frobenius norm loss are preferable. We repeat the entire procedure with $n=80$ and $n=200$, while keeping $n^*=40$ fixed, to examine the effect of increasing information through larger sample sizes. The results are reported in Figure \ref{fig:simstudy} which we discuss below. 

The first row in Figure \ref{fig:simstudy} shows the results for $\bm{B}$ for $n=40, 80$, and $200$, respectively. For $n=40$,
all four methods using a non-diagonal $\bm{\Sigma}$ face issues due to collinearity to some extent, as reflected in the inaccurate estimation of $\bm{B}$. For $n=80$, 
all the methods except dhs show an improvement. When $n=200$, all the methods show some improvement
with mssl and mbsp overcoming the problem altogether. The middle panel of the three plots shows that when the methods fail to estimate $\bm{B}$ accurately, they typically also fail to estimate $\bm{\Sigma}$ well. The third panel shows that the inaccuracy in the estimation of $\bm{B}$ can be large enough to affect the predictive performance of the methods. Among the four methods with a non-diagonal $\bm{\Sigma}$, we find that mbsp is least affected  by collinearity in the design matrix. However, even mbsp can be affected sometimes when $n=40$, as evident from Figure \ref{fig:simstudy}.


\section{NIR Spectroscopy Data} \label{Sec:real}
We apply the methods described in the previous section to a dataset containing measurements from near-infrared (NIR) spectroscopy. The goal of the experiment 
is to examine whether the composition of unbaked biscuit dough pieces can be determined by NIR spectroscopy. The $q=4$ response variables are percentages of fat, sucrose, dry flour, and water content in the biscuit dough pieces.
This dataset is called {\tt cookies} and is available as part of the R package 
{\tt fdm2id} \citep{fdm2id}. The dataset is known to have two outliers, which we removed before doing any analysis. The R package contains a training dataset containing $n=39$ samples, which we use to estimate the parameters. We use the provided test dataset with $n^*=31$ samples for prediction. For each piece of dough, $700$ measurements using NIR spectroscopy are available, which are the predictors. We select $p = 50$ predictors for our analyses to highlight the challenges posed by collinearity, as these predictor variables exhibit extremely high pairwise correlations (0.99 or greater).

We run the four MCMC based methods for 50,000 iterations including a burn-in of 100, and mssl is run with the maximum number of iterations set to 50,000. Posterior means are used for the estimation of $\bm{B}$ for the MCMC based methods and the corresponding posterior mode is used
for mssl. These estimates are used to predict the $n^*=31$ samples in the $n^* \times q$ test data matrix of response variables. The estimates of MCMC standard errors for mbsp are much larger than those for the other methods, so we re-run it for 2 million iterations. Although the standard errors remain somewhat larger than those of the other methods, we find the
predictive performance of mbsp did not change much between the 50,000 and the 2 million iterations runs. We report the results from the 2 million iterations run of mbsp. The values of Frobenius norm loss for two-step, dss, mssl, mbsp, and dhs are $20.521, 22.699, 30.710, 
18.206$, and $29.420$ respectively. To obtain a better understanding behind the higher values for mssl and dhs, we calculate the root mean squared error (RMSE) for prediction for 
each of the $q$ response variables, which are reported in Table \ref{tab:cookie}. Here mbsp is the best method, based on both the Frobenius norm loss and the prediction RMSE for each response variable in Table \ref{tab:cookie}. 
This demonstrates that modeling the dependency in the errors can be beneficial when there is sufficient information in the data, even in the presence of high collinearity among the predictors. However, since this would not be known in advance, performing the two-step analysis serves as a useful precaution.

\begin{table}[htb]
\centering
\begin{tabular}{crrrr}
  \hline
 Method & Fat & Sucrose & Flour & Water \\ 
  \hline
two-step & 1.527 & 2.366 & 1.483 & 0.820 \\ 
  dss & 1.527 & 2.659 & 1.746 & 0.872 \\ 
  mssl & 1.526 & 3.878 & 2.359 & 1.116 \\ 
  mbsp & 1.490 & 2.018 & 1.268 & 0.775 \\ 
  dhs & 1.538 & 3.598 & 2.410 & 1.035 \\ 
   \hline
   \end{tabular}
  \caption{RMSE for prediction of the 4 response variables: percentages of fat, sucrose, flour, and water in biscuit dough pieces, based on the {\tt cookies} test dataset.}
     \label{tab:cookie}
\end{table}

\begin{figure*}[htb]
    \centering
 \includegraphics[width=3.2in]{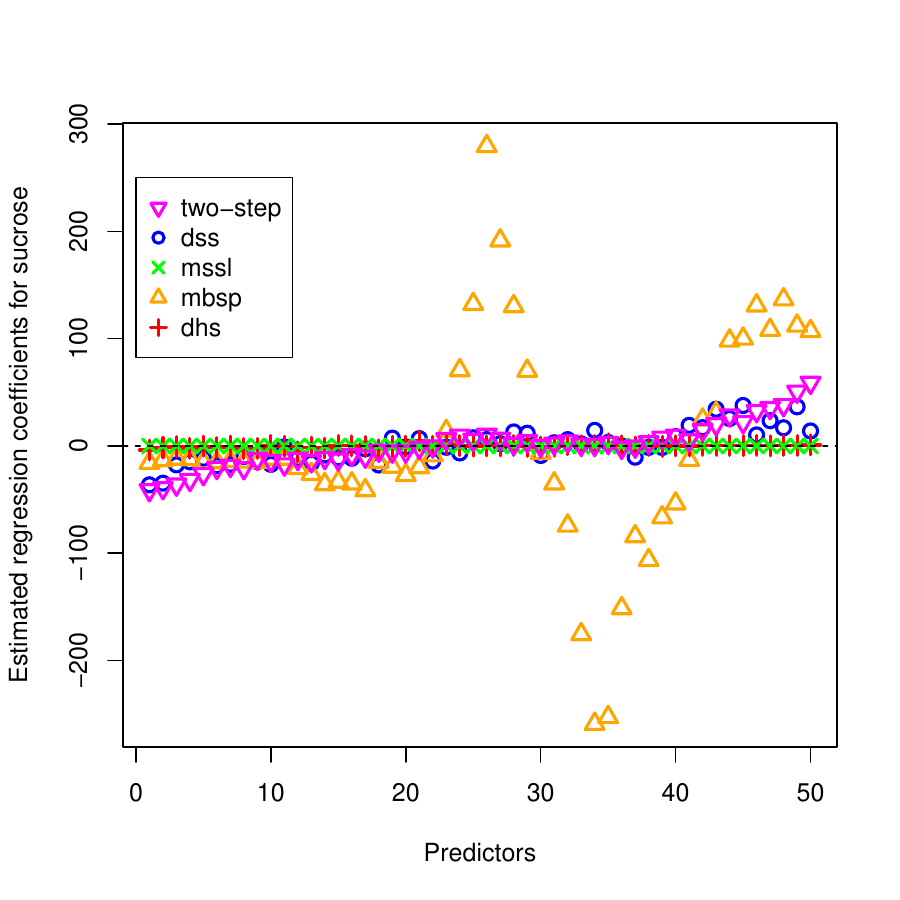} 
    \includegraphics[width=3.2in]{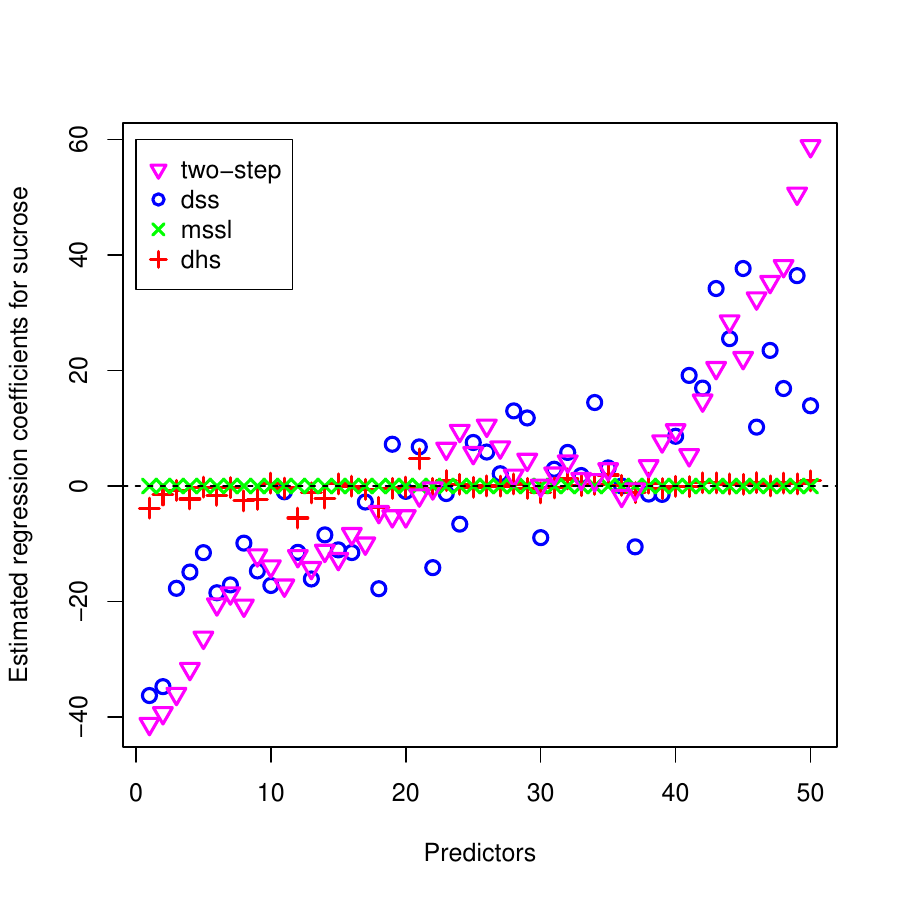}
    \caption{Estimated regression coefficients for the second response variable: percentage of sucrose in biscuit dough pieces, based on the {\tt cookies} dataset.}
      \label{fig:cookie}
\end{figure*}

The RMSE values for predicting the percentage of sucrose show one of the largest differences between the methods. So we plot the corresponding estimated regression coefficients (on the original scale after unstandardizing) for all methods, which correspond to the second column of the $\bm{B}$ matrix, in the left panel of Figure \ref{fig:cookie}. 
Since the regression coefficients from mbsp are much larger in magnitude than those from the other methods, we display the remaining four methods separately in the right panel of Figure \ref{fig:cookie} for clearer visualization. The plot in the right panel of Figure \ref{fig:cookie} shows that the mssl coefficients are all estimated to be zero, while the dhs estimates show some non-zero values but are very small. 
This suggests that, for this dataset, both mssl and dhs suffer from the effects of collinearity described earlier, with coefficients shrunk too aggressively toward zero, leading to higher prediction RMSE.

Since this is a real dataset, we do not know the true regression coefficients. The relatively large regression coefficients observed under mbsp raise the question of whether such magnitudes are truly necessary, or whether they reflect the influence of the horseshoe prior’s Cauchy-like heavy tails in the presence of collinearity, potentially resulting in inflated estimates.
 Whether more regularization through lighter tailed priors \citep{Ghos:Li:Mitr:2018, Ghos:2019} is suitable for this scenario is an open question. Under heavy-tailed posteriors, posterior medians are often considered more appropriate than posterior means. While this may generally be the case, it does not appear to hold under high collinearity in the design matrix. For this dataset, all methods shrink most regression coefficients to zero when using posterior medians, resulting in substantially worse predictive performance. This is not that surprising, as it is well-known that under high collinearity, the marginal inclusion probabilities of correlated predictors tend to be shrunk \citep{Ghos:Ghat:2015, Ghos:2015}, which can affect the median probability model of \citet{Barb:Berg:2004}. While using posterior median estimates of regression coefficients under Bayesian variable selection priors is not equivalent to using the median probability model, both approaches can be adversely affected by high collinearity for similar reasons, occasionally resulting in coefficients being shrunk to zero.

\section{Discussion} \label{Sec:Disc}
In this article, we examined the effect of collinearity among the predictors in Bayesian multivariate regression models with variable selection priors. While multivariate regression models that account for the dependency in the errors with a non-diagonal covariance matrix often improve the estimation of regression coefficients, our findings highlight the importance of exercising caution when collinearity is present. When collinearity is combined with a relatively small sample size and weak signals, there is less information in the data about all parameters. In such weakly identifiable scenarios, models with shrinkage priors and non-diagonal covariance matrices may become too complex to reliably estimate all parameters simultaneously, potentially leading to poor estimation and prediction. 

Our results indicate that performing an additional analysis using separate regression models for each response variable may help mitigate this issue. Among the Bayesian variable selection methods that we considered that use a non-diagonal covariance matrix, mbsp appeared to be the most robust to high correlation in the predictors. Since the posterior distributions are not available in closed form, it is difficult to pinpoint the exact reason for its superior
performance under collinearity. Its main difference from the other methods is that its prior on $\bm{B}$
depends on $\bm{\Sigma}$, whereas in all the other methods,
the elements of $\bm{B}$ are independent or conditionally independent given certain hyperparameters, and do not depend on $\bm{\Sigma}$. Additionally, mbsp uses a fixed global shrinkage parameter, unlike dhs, which places a prior on it. It is possible that the constraints imposed by the prior in mbsp improve identifiability of the parameters under collinearity, contributing to its stronger performance in such settings.

In this paper, we used discrete uniform priors
for the model space component of the spike-and-slab priors in the two-step approach. For a larger number of 
predictors, a Beta prior on the inclusion probability 
may be more appropriate. Since the identifiability issue is related to modeling the dependency in the errors across the response variables, the two-step approach
models each response variable using a  separate regression model, without borrowing any information across the response variables. An interesting direction for future work is to investigate whether borrowing of information through the model space prior \citep{Li:Ghos:Vill:2023:tas}
or the prior on regression coefficients \citep{Rich:Bott:Rose:2010, Das:Dey:Pete:Chak:2025} can be incorporated while maintaining a diagonal covariance structure, thereby mitigating the effects of collinearity.

\begin{center}
\Large{\textbf{Acknowledgments}}
\end{center}
    A part of this work was completed by Xun Li in her PhD thesis under the supervision of Joyee Ghosh.

\bibliography{refs.bib}   

\end{document}